# The Invariant Set Postulate:
# A New Geometric Framework for the Foundations of Quantum Theory and the Role Played by Gravity

T.N.Palmer
ECMWF, UK
tim.palmer@ecmwf.int
29 June 2009

**Abstract**

A new law of physics is proposed, defined on the cosmological scale but with significant implications for the microscale. Motivated by nonlinear dynamical systems theory and black-hole thermodynamics, the Invariant Set Postulate proposes that cosmological states of physical reality belong to a non-computable fractal state-space geometry $I$, invariant under the action of some subordinate deterministic causal dynamics $D_I$. An exploratory analysis is made of a possible causal realistic framework for quantum physics, based on key properties of $I$. For example, sparseness is used to relate generic counterfactual states to points $p \notin I$ of unreality, thus providing a geometric basis for the essential contextuality of quantum physics and the role of the abstract Hilbert Space in quantum theory. Also, self-similarity, described in a symbolic setting, provides a possible "realistic" perspective on the essential role of complex numbers and quaternions in quantum theory. A new interpretation is given to the standard "mysteries" of quantum theory: superposition, measurement, nonlocality, emergence of classicality and so on. It is proposed that heterogeneities in the fractal geometry of $I$ are manifestations of the phenomenon of gravity. Since quantum theory is inherently blind to the existence of such state-space geometries, the analysis here suggests that attempts to formulate unified theories of physics within a conventional quantum-theoretic framework are misguided, and that a successful quantum theory of gravity should unify the causal non-Euclidean geometry of space time with the atemporal fractal geometry of state space.

*"The task is not to make sense of the quantum axioms by heaping more structure, more definitions, more science fiction imagery on top of them, but to throw them away wholesale and start afresh. We should be relentless in asking ourselves: From what deep physical principles might we derive this exquisite structure? These principles should be crisp, they should be compelling. They should stir the soul."* Chris Fuchs, in Gilder (2008)

## 1 Introduction

The purpose of this paper is to propose a new geometric law of physics about the nature of physical reality on the cosmological scale, and to use it to perform an exploratory analysis of a possible causal realistic framework for quantum physics. Resulting from this analysis, new proposals are made about the role of gravity in quantum physics.

The Invariant Set Postulate is framed in terms of invariance, a concept that forms the very bedrock of physics, and conjectures that states of physical reality are defined by a fractal geometry $I$, embedded in state space and invariant under the action of some subordinate causal dynamics $D_I$. The postulate is motivated by two concepts that would not have been known to the founding fathers of quantum theory: the generic existence of invariant fractal subsets of state space for certain nonlinear dynamical systems, and the notion that the irreversible laws of thermodynamics are fundamental rather than phenomenological in describing the physics of extreme gravitational systems.

The notion that fractals may play a role in fundamental physics is not itself new. However, whilst earlier studies have focussed on the concept that space-time itself may be fractal (Ord, 1983; Nottale, 1984; El Naschie, 2004), the proposal in this paper concerns the ontological significance of fractal geometry in state space.

Although quantum theory is unsurpassed in terms of its agreement with experimental data, it is suggested that the Invariant Set Postulate provides a geometric framework for a deeper understanding of the foundations of quantum physics than can be provided by quantum theory itself. For example, as discussed in the body of this paper, the new perspective appears to reconcile many of Einstein's views about the incompleteness of quantum theory, with those of the standard Copenhagen Interpretation which emphasises the essential role of the observer in defining the very concept of reality.

Section 2 provides physical motivation for the Invariant Set Postulate as a new ontological postulate in physics. Section 2.1 reviews relevant aspects of nonlinear dynamical systems theory, whilst Section 2.2 discusses the state-space flow associated with the contents of the idealised "Hawking Box", a massive container holding enough matter to form one or more black holes. Here it is argued that the dynamical evolution of small volumes in the state space of the Hawking Box will plausibly asymptote to the types of zero-volume invariant sets discussed in Section 2.1. These ideas are combined to formulate the Invariant Set Postulate, discussed in Section 3.

Section 4, the heart of the paper, discusses quantum theory assuming the Invariant Set Postulate. On the invariant set, the quantum-theoretic state is straightforwardly interpreted as defining a probability with well-defined sample space (Section 4.1). In terms of this, the sparseness of fractal invariant sets in state space provides a novel explanation why any realistic interpretation of quantum theory must be contextual, as required by the Bell-Kocken-Specker theorem (Section 4.2); essentially, dynamically unconstrained counterfactual states are associated with points of unreality $p \notin I$. This in turn leads to a new perspective on the necessity for the abstract Hilbert space formulation of quantum theory, a theory which does not distinguish between real and counterfactual states; the quantum-theoretic state of a quantum sub-system is defined for all conceivable measurement that might be made on it. In this perspective, an abstract "probability state" is associated with neighbourhoods of of state space not belonging to $I$, but now there is underlying sample space and the mathematical properties of this abstract probability state are defined merely by appropriate algebraic properties, here that of the vector space (Section 4.3). Since $I$ is non-computable, it is algorithmically undecidable and hence irreducibly uncertain, whether or not a quantum state vector can be associated with an underlying sample space on the invariant set, or not.

A key property of the quantum Hilbert Space is that it is complex. A realistic interpretation of complex numbers and quaternions is provided in Section 4.4 in terms of permutations on bit strings that correspond to bivalent sample spaces of trajectory labels on $I$. It is suggested that state-space magnification of $I$'s self-similar structure by $D_I$'s positive-exponent Lyapunov vectors provides the geometric basis for periodicity in these permutations, thus leading to a novel explanation for the notion of quantum coherence.

Combining these perspectives, a comprehensible account of many of the standard "mysteries" of quantum theory is outlined in Section 5, including new speculations concerning the role of gravity in quantum physics.

Concluding remarks are given in Section 6.

Throughout this paper, when the word "state" is written without further qualification, it is meant in its conventional classical meaning. Similarly the phrase "state space" is meant in its classical sense (where it generalises the notion of phase space eg of an N-particle system). When the notion of "state" is meant its quantum mechanical sense, ie as an element of a Hilbert Space, then the word will be clearly qualified eg by referring to the "quantum-theoretic notion of state" or, occasionally, as the "quantum state vector".

## 2 Motivations for the Invariant Set Postulate

The remarks in this section provide motivation for the Invariant Set Postulate, to be defined in Section 3. There are two different aspects to this motivation.

**2.1 Dynamically Invariant Sets**

Since one of the objectives of this paper is to develop a new perspective on the role of gravity in quantum theory, it is appropriate to begin discussion with mention of the seminal work of Poincaré, and Birkhoff who found that the motion of as few as three gravitationally-bound Newtonian particles is chaotic, ie with aperiodic evolution in a bounded domain of state space. By considering entire cosmologies such as the Mixmaster solution (Misner et al, 1973), it appears general relativity also admits chaotic space-time dynamics. The use of geometric concepts, based on the type of invariant set to be considered below, appears to be essential for a relativistically-invariant definition of chaos (Cornish, 1997).

Over a half century after Poincaré, Lorenz (1963) proposed a very different type of chaotic motion, associated with forced nonlinear dissipative dynamical systems $\dot{\mathbf{X}} = \mathbf{f}(\mathbf{X})$. In contrast with Hamiltonian systems, the states $\mathbf{X}(t)$ of such systems evolve asymptotically to fractionally-dimensioned (fractal) attractors. If $\mathbf{X}$ is initialised on an attractor, $\mathbf{X}$ stays on it forever; the attractor is a dynamically-invariant subset of state space.

Fractal attractors reveal some of the most beguiling of geometries known to physics, and form the basis of discussion in this paper. However, the dynamical systems which generate these geometries are usually considered phenomenological rather than fundamental, since they are explicitly dissipative. In Section 2.2, a inherently relativistic reason for supposing these geometries to be fundamental will be proposed.

Such a dynamically-invariant attractor can be formed from the asymptotic evolution of a volume $V(t)$ in state space under the action of the dissipative ($\nabla \cdot \dot{\mathbf{X}} < 0$) dynamics. In the case of the Lorenz (1963) system, $\nabla \cdot \dot{\mathbf{X}}$ is constant over state space, in other systems (eg Rössler, 1976) $\nabla \cdot \dot{\mathbf{X}}$ is flow dependent but negative overall. As discussed in Section 2.2, systems of the latter type are envisaged as describing a dynamically-invariant set of spatial cosmologies $\mathbf{X}$, where regions $\nabla \cdot \dot{\mathbf{X}} < 0$ describe states of the universe containing black holes.

The following properties of chaotic invariant sets will be needed in the discussions in Sections 3-5.

The invariant sets under consideration have zero measure and are nowhere dense in the Euclidean state space in which they are embedded. As such, a perturbation which is random with respect to the continuum measure of state space will almost certainly map a point on the invariant set, off it. As an example, consider the classical Cantor set. A point on the Cantor set can be represented on the interval [0,1] by an exceptional fraction $n_C$ with no digit "1" in its base-3 representation. By contrast a random perturbation can be represented on the interval by a normal (Hardy and Wright, 1979) fraction $n_r$ whose base-3 representation has, for example, equal frequencies of the digits "0", "1" and "2". Almost certainly the "perturbed" number $n_C + n_r$ will not be normal, (eg it will contain "1" digits), and hence cannot represent a point on the Cantor set. The sparseness of fractal invariant sets is central to a new perspective on counterfactual reasoning, discussed in Section 4.

Using the *p*-adic metric (Khrennikov, 1997), these invariant sets can be shown to be metric spaces and it is therefore possible to talk about neighbourhoods on the invariant set. Interestingly, p-adic numbers have been studied in the context of mathematical physics (eg Alberverio and Khrennikov, 1998). It remains to be seen whether ideas in this paper can be linked quantitatively to such studies.

Fractal invariant sets are non-algorithmic. For example, Dube (1993) has shown that the invariant sets of iterated function systems emulate the non-halting states of Turing machines, and undecidable problems in the classical theory of computation have a corresponding geometric interpretation. For example, the Post Correspondence Problem is equivalent to asking whether a given line intersects the invariant set of an iterated function system. More generally, Blum et al (1998) have shown that if an invariant set has fractal dimension, it cannot be a halting set. Below it is proposed that the undecidability of these invariant sets provides the fundamental basis for quantum uncertainty, ie uncertainty as to whether or not a specific quantum state vector can be associated with an underlying sample space. Penrose (1989) has given arguments for why non-computability might lie at the heart of fundamental physics.

Invariant sets can be constructed from time series of experimental data, using the Takens Embedding Theorem (Takens, 1981). Suppose the time series of some component of **X** is measured every $\tau$ units of time. Then the invariant set can be reconstructed from a sufficiently long time series of this component. An important conceptual point relevant to the discussion on free variables in Section 5, is that a sufficiently long time series of even an energetically unimportant or otherwise seemingly irrelevant component of **X** can be used to construct the entire invariant set.

One key technique to represent the evolution of states on an invariant set is that of symbolic dynamics (Lind and Marcus, 1995; beim Graben and Atmanspacher, 2006). Consider a bivalent partition $\Pi$ of $I$ based on disjoint subsets $A, B \subset I$ such that all points of $I$ either belong to $A$ or to $B$, and a sequence eg *.AABABB...* where the *n*th member of the sequence represents the subset in which **X** belongs at the *n*th iterate of the dynamical evolution operator. Then the same sequence *.AABABB..* can symbolically represent **X** at the first iterate, and dynamical evolution is effected by a simple shift of the sequence one place to the left, relative to the radix point. In the case of so-called generating partitions, this symbolic representation is homeomorphic to the original dynamics. For the situations considered here, "*A*" and "*B*" will label trajectories in the basin of attraction of relatively stable quasi-stationary regions of the invariant set, corresponding to "measurement outcomes". In general, such "*A*" and "*B*" trajectories will be intertwined with respect to one another in neighbourhoods on the invariant set (Palmer, 1995). In this case, the bit string *.AABABB...* defines a sample space of trajectory segments. As discussed in Section 4, permutations of such bit strings can be defined which are identical to the unit quaternions, themselves linked to the algebra of quantum spin (Palmer, 2004).

An important property of fractal sets is self similarity. Readers may be familiar with animations which zoom into the Mandelbrot set revealing periodicity in the intricate fractal structure. Lapidus and van Frankenhuis (2006) in their study of fractal geometry were able to exploit the self similarity of fractal strings to assign them with a complex dimension, defined from the poles of a corresponding zeta function. Below the positive-exponent Lyapunov vectors of the dynamics associated with the invariant set are considered to provide the dynamic "zoom", thereby generating periodicity in coarse-grain probability measures with respect to the intertwined partition.

As discussed in Section 4, this may provide a dynamical basis for the oscillatory nature of the quantum state vector, from the perspective of the Invariant Set Postulate.

A key conceptual component of the perspective pursued below, is that the geometry of the invariant set should be considered as more primitive than the differential equations whose asymptotic behaviour generates the invariant set. This is not the normal perspective used in dynamical systems theory where the difference or differential equations are primary. In this respect it is worth commenting on some of the global geometric and topological approaches to defining fractal invariant sets. In studying the structure of chaotic attractors in 3-dimensional state space, Birman and Williams (1983) focussed on the knottedness of the associated unstable periodic orbits. For example, the knots describing the Lorenz (1963) attractor are prime, fibered with non-negative signature. In higher dimensions, studies have been made characterising fractal invariant sets using simplicial spaces and corresponding homology invariants (Sciamarella and Mindlin, 1999). In this respect, it is worth noting that the very subject of topology arose in significant measure from Poincaré's attempts to understand and characterise the behaviour of the gravitational 3-body problem, in the absence of any general analytic solution to the problem.

**2.2 The Hawking Box**

The last section began with Poincaré's work on Newtonian gravitational systems. Here we discuss the role of an extreme non-Newtonian gravitational system which may be pivotal in generating the fractal properties of the invariant set studied in this paper.

Conventional physics is formulated in terms of Hamiltonian dynamics, and the state-space flow is consequently incompressible: $\nabla \cdot \dot{\mathbf{X}} = 0$. However, consider a compact system big enough that a black hole could potentially form from the collapse of matter comprising the system. How can we characterise the asymptotic state space flow of such a system? Is it Hamiltonian?

This has been the topic of considerable debate, especially between two of the leading experts in gravitation theory (Hawking and Penrose, 2000). The debate hinges around a thought experiment in which some vast but gravitationally isolated system is placed in a hypothetical box with reflecting walls - the Hawking Box. Indeed we could imagine the Hawking Box without boundary, since it is observationally possible that the observed universe may have compact spatial topology.

What is the asymptotic state-space flow of the Hawking Box?

Two of the most important results in 20th Century theoretical physics are relevant here: the proof that space-time singularities formed by gravitational collapse are generic (Penrose, 1965), and the quantum field theoretic calculation that black holes formed as a result of such gravitational collapse have precise thermodynamic properties (Hawking, 1975). The second law of black-hole thermodynamics describes the loss of information as matter collapses to a black hole, and subsequently evaporates as thermal radiation.

As a result of the second law, Penrose (eg in Hawing and Penrose, 2000) argues that the state space flow of the Hawking Box must be convergent $\nabla \cdot \dot{\mathbf{X}} < 0$ in regions of state space containing one or more black holes. Penrose also argues that shrinking of state-space volumes contradicts Liouville's theorem and concludes that there must therefore be compensating regions of state-space divergence, thus motivating Penrose's (1989) gravitationally-induced Objective Reduction mechanism for quantum state-vector collapse.

A thorough review of the Liouville theorem has been given recently by Ehrendorfer (2006). Liouville's original paper (Liouville 1838) concerns a result on the material derivative of the Jacobian of the mapping between a solution of a differential equation and its initial state, and is not specific to Hamiltonian systems. In a simple form, the Liouville equation for the probability density function $\rho$ of the state vector $\mathbf{X}$ is

$$\frac{\partial \rho}{\partial t} + \nabla \cdot (\rho \dot{\mathbf{X}}) = 0 \tag{1.1}$$

reminiscent of the Newtonian mass continuity equation in physical space. If the system is Hamiltonian so that $\nabla \cdot \dot{\mathbf{X}} = 0$, then co-moving volumes are conserved in state-space. On the other hand, even for systems where the state-space flow is compressible, the Liouville equation (1.1) guarantees conservation of probability. That is to say, Liouville's theorem in its general (ie non-Hamiltonian) form does not require co-moving volumes in state space to be conserved. If the existence of flow convergence due to black hole thermodynamics is being assumed, there is no requirement to use the more restrictive volume-preserving form of the Liouville equation. As such, Liouville's theorem does not itself imply the need for the Objective Reduction mechanism.

Hence, suppose state-space volumes shrink in those parts of state space containing black holes, but without any corresponding divergence due to Objective Reduction. What are the consequences? The situation is analogous to the dissipative dynamical systems' approach in the previous section with state-dependent $\nabla \cdot \dot{\mathbf{X}}$. It can therefore be supposed that volumes $V(t)$ will shrink to zero and asymptote onto one of the following: a fixed point, a limit cycle, or a fractal attractor. Since fractal attractors are generic, we will assume henceforth that the zero-volume asymptotic limit is a fractal set.

Clearly it will take an infinite time for a generic volume in the state space of the Hawking Box to evolve precisely onto its invariant set. However, treating the Hawking Box as a model of the universe, we ask in the next section whether there might exist a hitherto unknown law of physics, defined on the cosmological-scale, which postulates that the universe as a whole lies precisely on its invariant set, $I$. As a dynamically-invariant set, then if the universe lies on $I$ now, then it always has lain on it, and always will lie on it. Thus this putative law of physics, named below the "Invariant Set Postulate", is best thought of as an atemporal geometric constraint in state space. Although defined on the cosmological scale, the Invariant Set Postulate will be shown to have fundamental ontological implications for physics on the microscale.

## 3 The Invariant Set Postulate

The Invariant Set Postulate posits the existence of a fractionally-dimensioned subset $I$ of the state space of the physical world (ie the universe as a whole). $I$ is an invariant set for some presumed-causal (ie relativistic) deterministic dynamical system $D_I$; points on $I$, hereafter referred to as world states, remain on $I$ under the action of $D_I$. World states of physical reality are those, and only those, lying precisely on $I$.

It is conceptually important to view $I$ as more primitive than $D_I$. Given $I$, $D_I(t)$ maps some point $p \in I$ a parameter distance $t$ along a trajectory of $I$. Crucially, $D_I$ is undefined at points $\notin I$. This contrasts with the more familiar situation where a dynamical system is defined by differential or difference equations $D$, which asymptotically generates an invariant set $I_D$. In this latter, classical, situation, $D$ is defined at all points of state space.

If states of physical reality necessarily lie on $I$, then points $p \notin I$ in state space are to be considered literally "unreal". In a hypothetical "oracle" theory of physics which (non-computability notwithstanding) had perfect knowledge of $I$, these points of unreality would be an irrelevance. However, for practically-relevant theories (such as quantum theory and any algorithmic extension) the intricate structure of $I$ is unknown and these points of unreality cannot be ignored. We return to this in Section 4.3 where one of the key questions considered is how to represent quantum-theoretic states in a mathematically-consistent way for such points of unreality.

The search for an atemporal description of physics has been long-standing (eg Barbour, 1999; Price, 1996). The Invariant Set Postulate provides support to this search: treating the geometry of the invariant set as primitive, introduces a fundamentally atemporal perspective into the formulation of basic physics. Such a perspective is absent in classical physics, in which differential equations are considered primitive. This has ramifications for a new perspective on the emergence of classicality in quantum physics, discussed further in Section 5.

Since no classical theory requires states to be constrained to invariant sets, even when the theory supports such sets, the proposal here does not constitute a return to a "classical physics" formulation of quantum theory.

## 4 Quantum Theory and the Invariant Set Postulate

In this section the perspective brought to quantum theory by the Invariant Set Postulate is discussed.

### 4.1 Probability on the Invariant Set

Quantum-theoretic descriptions of sub-system states fall under the general definition provided by Hardy (2004): the state of a system is defined to be that thing represented by any mathematical object which can be used to predict the probability associated with every measurement that may be performed on the system.

Quantum-theoretic states are defined in terms of an abstract complex Hilbert Space. Hence, if $|A\rangle$ and $|B\rangle$ are quantum states, then so is $|\psi\rangle = \alpha|A\rangle + \beta|B\rangle$, where $\alpha, \beta \in \mathbb{C}$. It is, of course, one of the great mysteries of quantum theory (some would say the central mystery) as to the physical reality of the superposed state. The Invariant Set Postulate provides a simple answer to this: on the invariant set, and only on it, $|\psi\rangle = \alpha|A\rangle + \beta|B\rangle$ can be interpreted as defining a probability mixture of two discrete alternatives based on a well-defined associated sample space. The issue of why $\alpha, \beta$ must be complex rather than real numbers, is discussed in Section 4.4.

For concreteness, suppose that Bob measures the spin of a spin-1/2 particle as "up" relative to a prepared Stern-Gerlach apparatus. How can this be described in the context of the Invariant Set Postulate? If the particle detector in the "up" beam fires at time $t_1$, label the world state at $t_1$ by the letter "*A*". Now track the world state trajectory backwards on the invariant set to an earlier time $t_0$ before the particle entered the Stern Gerlach apparatus, and now label the back trajectory from $t_0$ to $t_1$ by the letter "*A*". Consider a neighbouring trajectory at $t_0$, also on the invariant set. If this trajectory passes through a region of instability between $t_0$ and $t_1$ (ie where neighbouring trajectories diverge), then it may conceivably evolve to a world state at $t_1$, in a distinctly different quasi-stationary region of the invariant set corresponding to a situation where the particle detector in the "down" beam fires. Such neighbouring trajectories will be labelled "*B*".

If the neighbourhood of "*A*" at time $t_1$ is essentially a quasi-stationary region of relative stability of $I$, then all neighbouring trajectories in this region can be labelled "*A*". Similarly, all trajectory segments in the neighbourhood of "*B*" at time $t_1$ can similarly be labelled "*B*". By contrast, a neighbourhood of "*A*" at time $t_0$ will comprise a mixture of both "*A*" and "*B*" trajectories, indeed in this region the "*A*" and "*B*" trajectories will be intertwined with respect to one another on the invariant set (Palmer, 1995), describable as a probability mixture. Following the order-of-magnitude analysis of Penrose (1989) we associated this heterogeneity of relative stability on $I$ as being gravitationally induced, and the evolution of the world state towards a relatively stable quasi-stationary end point, as what we would experimentally refer to as a "measurement outcome".

From this point of view, it can be supposed that the quantum-theoretic state $|\psi\rangle = \alpha|A\rangle + \beta|B\rangle$ defines the probability that a randomly chosen trajectory segment in a neighbourhood of some $p \in I$ is an "*A*" trajectory or a "*B*" trajectory. Here the word "random" is defined with respect to an appropriate metric on the invariant set, see Section 2.1. Since Planck's constant has the dimensions of phase-space area, we can use $\sqrt{\hbar}$ to fix the dimensional size of this neighbourhood. At $t_0$, "upstream" of a region of instability, we can expect both $\alpha$ and $\beta$ to be non-zero, whilst at $t_1$, in a region of stability, either $\alpha$ or $\beta$ will be equal to zero. This evolution is what otherwise could be referred to as a "collapse" of the quantum state.

In the laboratory, particle spin statistics are derived, for example, from a sample of measurements performed on a set of identically prepared particles.

These measurements and their "up/down" outcomes are properties of a single "fiducial" trajectory segment on the invariant set. By contrast, the notion of probability described above is based on a sample of trajectories in the neighbourhood of the fiducial trajectory. Hence it must be assumed that the statistical ensemble comprising neighbouring trajectories each with some specific $(A, B)$ label, can be identified with a statistical ensemble of measurement outcomes specific to the fiducial trajectory. This issue is discussed further in Section 4.4.

**4.2 Contextuality and the Ontology of Counterfactual States**

The notion that there is some underlying causal deterministic process which generates the sample space from which the quantum-theoretic notion of state can be defined, appears to run counter to the Bell-Kochen-Specker theorem which clearly asserts that no such realistic interpretation of the quantum-theoretic state is possible. We discuss here why this theorem does not apply under the conditions of the Invariant Set Postulate.

In conventional hidden-variable theory, it is assumed that a preparation procedure creates a quantum sub-system with certain intrinsic properties, and that different measurements reveal different aspects of these properties. The Bell-Kochen-Specker theorem shows that such non-contextual models are inconsistent with quantum theory. On the other hand, by requiring the world state to lie on its sparse invariant set, such a sub-system's properties will not in fact be independent of the rest of the world state, and in particular of how the measurement device has been prepared. That is, the assumption that a sub-system necessarily has properties independent of the conceivable measurements made on it, may be wrong. The Invariant Set Postulate provides a physically-appealing basis for constructing a contextual hidden-variable theory of quantum physics.

To be more explicit, suppose Bob measures the spin of a spin-1/2 particle with his Stern-Gerlach apparatus oriented in the $z$ direction. What would the measurement outcome have been had he measured at some angle $\theta$ to the $z$ direction? To attempt to give meaning to this question, we imagine a counterfactual world where everything, including the spin-1/2 particle, is as it was in the real world, except that the orientation of the Stern-Gerlach apparatus, here represented by the variable $\theta$, has been changed. Can these counterfactual worlds be considered elements of physical reality?

Let $p \in I$ denote the world state at the time Bob measures the spin of the particle. Let $l_\theta$ denote the line in state space passing through $p$ and pointing in the state space direction where $\theta$ varies, but where the values of all other components of the world state stay fixed. The intersections of $l_\theta$ with $I$ at points other than $p$ define the set $\{\theta_C\}$ of counterfactual values for $\theta$ allowed by the Invariant Set Postulate (points of intersection are counterfactual states which are also states of physical reality). Given $\{\theta_C\}$, we can define a binary function $Sp(\theta_C) \in \{A, B\}$ such that, according to the dynamics, Bob would have measured "up" if $Sp(\theta_C) = A$, and spin "down" if $Sp(\theta_C) = B$.

However, the sparseness of $I$ suggests that counterfactual sets such as $\{\theta_C\}$ are in fact equal to the empty set. Consider, for example, a two dimensional set $I_2$, formed as the Cartesian product of two Cantor sets, whose $(x, y)$ coordinates are pairs of ternary fractions, each with the missing digit "1". Now consider a line $l$ passing through the origin and oriented randomly in the sense that the gradient $k$ of $l$ is a normal number. If $l$ were to intersect $I_2$ at a point with coordinates $(x_C, y_C) \neq (0, 0)$, then both $x_0$ and $y_0$ must be exceptional ternary fractions with missing digit "1". However, this is almost certainly not the case; since $k$ is normal, then a ternary expansion of $y_C = kx_C$ almost certainly contains the digit "1" even if the ternary expansion of $x_C$ does not contain it.

The emptiness of the counterfactual set, which we here assume, is consistent with the implications of the Bell-Kochen-Specker theorem, that there can be no non-contextual hidden-variable representations of individual quantum systems. If there were a non-contextual hidden variable model, one could vary any one of its parameters keeping fixed other variables corresponding to the rest of the world state, to obtain new elements of supposed physical reality.

**4.3 Towards the Abstract Hilbert Space**

If quantum theory could "see" the intricate structure of the invariant set, it would "know" whether a particular putative measurement orientation $\theta$ was counterfactual or not. However, since, by hypothesis, quantum theory is blind to the intricate structure of $I$, it is unable to discriminate between factual and counterfactual measurement preparations and therefore admits them all as theoretically valid. Hence the quantum-theoretic notion of state is defined on a quantum sub-system in preparation for any measurement that could conceivably be performed on it, irrespective of whether this measurement turns out to be real or counterfactual. This raises a fundamental question. If we interpret the quantum-theoretic notion of state in terms of a sample space defined by a $\sqrt{\hbar}$ neighbourhood on the invariant set, how are we to interpret the quantum-theoretic notion of state associated with counterfactual world states of unreality, not on the invariant set, where no corresponding sample space exists?

Consider the following analogy. The (rational) integers are rudimentary symbols of counting, used for example to express and compare the quantity of apples in piles of apples. As a consequence of being symbols of counting, these integers have certain algebraic properties: the sum, difference or product of two integers is a third. Based entirely on these algebraic properties, it is possible to extend the notion of integer to the Gaussian integers on the complex plane. These algebraic integers are <u>defined</u> by their algebraic properties, and no more have the primitive property of being symbols of counting; no piles of apples contain $1+2i$ apples! As long as we are not concerned whether an integer can be used to count piles of apples, then each point $p$ of a Cartesian grid in the complex plane defines an integer. On the other hand, if we are told, as a result of some empirical study, that the integer at $p$ describes the quantity of some particular pile of apples, then we can infer that $p$ must lie on the real axis of the complex plane!

This analogy is useful in arriving at the required generalisation of the quantum theoretic notion of state in a theory blind to the intricate structure of the invariant set.

Hence, when $p \in I$ (c.f. the real axis for the Gaussian integers), then $\alpha|A\rangle + \beta|B\rangle$ can be interpreted as a probability defined by some underlying sample space. However, when $p \notin I$ (c.f. the rest of the complex plane for the Gaussian integers), we define a probability-like state $\alpha|A\rangle + \beta|B\rangle$ from the algebraic properties of probability, ie in terms of the algebraic rules of vector spaces. Under such circumstances, $\alpha|A\rangle + \beta|B\rangle$ can no more be associated with any underling sample space. This "continuation off the invariant set" does not contradict Hardy's definition of state, since if $p \notin I$, then its points are not elements of physical reality, and hence cannot be subject to actual measurement.

It is worth discussing the corresponding situation in classical physics. A classical dynamical system is one defined by a set of deterministic differential equations. As such, there is no requirement in classical physics for states to lie on an invariant set, even if the differential equations support such a set. (Indeed, for systems which have a fractal invariant set, the probability that a state lies on it precisely is zero, and the invariant set is thus an ontological irrelevance.) As a result, for a classical system, every point in phase space is a point of "physical reality", and the counterfactual states discussed above are as much states of "physical reality" as are the real world states. Hence, the world of classical physics is perfectly non-contextual, and is not consistent with the Invariant Set Postulate. The question of how classicality can emerge from the Invariant Set Postulate is discussed in Section 5.6.

The following interpretation of the two dimensional Hilbert Space spanned by the vectors $|A\rangle$ and $|B\rangle$, eg in the context of Bob's spin measurements, emerges from the Invariant Set Postulate. At any time $t$ there corresponds a point in the Hilbert Space where $\alpha|A\rangle + \beta|B\rangle$ can be interpreted straightforwardly as a frequentist probability based on an underlying sample space of trajectory segments in a $\sqrt{\hbar}$ neighbourhood on the invariant set. However, since the invariant set and hence its underlying deterministic dynamics are themselves non-computable, it is algorithmically undecidable as to whether any given point in the Hilbert Space can be associated with such a sample space or not; as such, each point of the Hilbert Space is as likely to support an underlying sample space as any other. For points in the Hilbert Space which have no correspondence with a sample space on the invariant set, $\alpha|A\rangle + \beta|B\rangle$ must be considered an abstract mathematical quantity defined purely in terms of the algebraic rules governing a vector space.

Consistent with the rather straightforward probabilistic interpretation of the quantum-theoretic notion of state on the invariant set, it is reasonable to suppose that, on the invariant set, the Schrödinger equation is itself a Liouville equation for conservation of probability in regions where dynamical evolution is Hamiltonian (ie in regions not associated with black holes). Since quantum theory is blind to the intricate structure of the invariant set, the quantum-theoretic Schrödinger equation must be formulated in abstract Hilbert space form, ie in terms of unitary evolution, using algebraic properties of probability without reference to an underlying sample space.

One algebraic property inherited from the Schrödinger equation's interpretation as a Liouville equation on the invariant set, is linearity: as an equation for conservation of probability, the Liouville equation, cf equation (1.1), is always linear, even when the underlying dynamics $\dot{\mathbf{X}} = \mathbf{f}(\mathbf{X})$ are strongly nonlinear. This suggests that attempts to add nonlinear terms (deterministic or stochastic) to the Schrödinger equation eg during measurement, are misguided.

**4.4 Role of the Complex Numbers**

The quantum-theoretic state is an element of a complex-number Hilbert Space, $\alpha, \beta \in \mathbb{C}$. The discussion so far has not so far touched on the essential role that the complex numbers play in quantum theory. We discuss in this Section a novel interpretation for the role of the complex numbers, in the context of the Invariant Set Postulate.

The use of complex numbers, ie $\alpha, \beta \in \mathbb{C}$ in $\alpha|A\rangle + \beta|B\rangle$ and more specifically the use of quaternions in representing the quantum-theoretic notion of spin, is an essential element of quantum theory. How can such complex numbers be represented in the "realistic" context of the Invariant Set Postulate? Let

$$S = (a_1, a_2, a_3, a_4, \ldots) \quad (2.1)$$

where $a_i \in \{A, B\}$ denote a sample space of "A" or "B" trajectories in a neighbourhood of the invariant set from which, as in Section 4.1, $\alpha|A\rangle + \beta|B\rangle$ is defined. Now let $\overline{A} = B$ and $\overline{B} = A$ and

$$-S = (\overline{a_1}, \overline{a_2}, \overline{a_3}, \overline{a_4}, \ldots) \quad (2.2)$$

so that $S$ and $-S$ denote two anti-correlated sample spaces. Writing

$$i(S) = (\overline{a_2}, a_1, \overline{a_4}, a_3, \ldots) \quad (2.3)$$

(with operator repetition on successive pairs of elements) then $i^2(S) = -S$, so that $i$ is a "square root of minus one", and $S$ and $i(S)$ are uncorrelated but nevertheless dependent sample spaces ( ie $i(S)$ depends on $S$). Similarly, writing

$$\begin{aligned} e_1(S) &= \{\overline{a_2}, a_1, a_4, \overline{a_3} \ldots\} \\ e_2(S) &= \{\overline{a_3}, \overline{a_4}, a_1, a_2 \ldots\} \\ e_3(S) &= \{a_4, \overline{a_3}, a_2, \overline{a_1} \ldots\} \end{aligned} \quad (2.4)$$

(with operator repetition on successive quadruplets of elements) then each of $e_1, e_2, e_3$ is also a "square root of minus one" and collectively they satisfy the rules of quaternionic multiplication, ie
.
$$e_1 \circ e_1(S) = e_2 \circ e_2(S) = e_3 \circ e_3(S) = e_1 \circ e_2 \circ e_3(S) = -S \quad (2.5)$$

It should be noted that the representations (2.4) are not unique, other sets of quaternionically-related strings can be defined from permutation operators acting on successive octuplets and so on; see Palmer, 2004. Using these more general representations, then strings $S_1'$, which are partially correlated both with $S$ and with, say, $e_1(S)$, are readily constructed.

As is well know, the Pauli spin matrices are themselves related to quaternions and exploit the (surjective) homomorphism between $SU(2)$ and $SO(3)$.

These permutation representations suggest the following symbolic proposal for the role of the complex numbers in the context of the Invariant Set Postulate. For neighbourhoods on the invariant set, where the quantum state $\alpha|A\rangle + \beta|B\rangle$ is associated with an underlying sample space represented by the binary sequence $S$, the Pauli operators have granular permutation-based representations (here $\sigma_z = ie_1$, $\sigma_y = ie_2$, $\sigma_x = -ie_3$), which map $S$ to deterministically related sample spaces eg $S_1'. S_2', S_3'....$ In this way, it is suggested that compound "entangled" quantum-theoretic states can be associated with multiple bit strings ie multiple binary sample spaces, which are deterministically related to one another. Each of the $S_1'. S_2', S_3'...$ represents a binary sample space of labels over a common set of neighbouring trajectories. Using the linkage between $SU(2)$ and $SO(3)$, these labels can be taken to refer to "yes/no" measurement outcomes associated with multiple orientations in physical 3-space.

Off the invariant set, the notion of "entanglement" can only be defined in the context of the abstract Hilbert Space, ie without any underlying sample spaces, consistent with the discussion in Section 4.3.

A key element of the geometry of fractal invariant sets is self similarity. As mentioned in Section 2.1, such self-similarity can be revealed by "zooming" into a static fractal set. In the case of a dynamic set, self similarity can be revealed by state-space amplification by the positive-exponent Lyapunov vectors. Consider a set of trajectories on the invariant set in the neighbourhood of the fiducial trajectory, and labelled by the bit string $S$. Through the action of the positive-exponent Lyapunov vectors, these trajectories will diverge from one another and eventually leave the neighbourhood. In their place will "appear" (relative to some scale defined by the appropriate metric on the invariant set, cf the resolution of the eye when viewing fractal sets) a replacement set of trajectories, previously existing on some previously small scales. Label this replacement set of trajectories by some new bit string $S'$. Self similarity can then be associated with an oscillation of the form $S \rightarrow S' \rightarrow S \cdots$. Using the quaternionic bit-string representations above, one can express corresponding self similar oscillations by the form

$$\cdots S \rightarrow e_1(S) \rightarrow -S \rightarrow -e_1(S) \rightarrow S \rightarrow \cdots \qquad (2.6)$$

of the phase-quadrature components, thus providing the basis for a "realistic" representation of phase evolution $\exp(i\omega t)$, as required by the Schrödinger equation.

This geometric self-similar structure must be manifest on the fiducial trajectory itself. Consistent with the discussion at the end of Section 4.1, it must be assumed that $S$, and the associated quaternionic mappings of $S$, can also represent an ensemble of labels of the fiducial trajectory segment relative to different state-space directions.

These relationships between the complex numbers on the one hand, and the symbolic structure of $I$ on the other, provide profound constraints on the topology of the invariant set. These constraints rule out the Lorenz or Rossler attractors, or other standard dynamical systems models, as viable "toy" models of the invariant set of the universe, even though such invariant sets are themselves fractal. Given the non-computability of fractal invariant sets, the analysis here suggests that the most likely approach to finding a robust mathematical representation of the invariant set of the universe is through the symbolic approach; as discussed in Section 2.1, this approach can define the invariant set to topological equivalence. Indeed, as remarked by Bohr himself, the very fact that a quantum-theoretic state has the form $\alpha|A\rangle + \beta|B\rangle$, suggests that quantum theory is itself profoundly symbolic.

## 5 An Invariant-Set Perspective on the "Mysteries" of Quantum Theory

In Section 4, a new approach to the foundations of quantum physics has been outlined based on an ontological postulate defined on the cosmological scale. The approach is not yet in the form of a rigorous mathematical theory which can somehow "subsume" standard quantum theory. Nevertheless, it is appropriate to discuss the implications of this approach in helping to make sense of the standard "mysteries" of quantum physics, in terms of ideas we can comprehend. Hence, in this section, the ideas developed in Section 4 are put to work.

**5.1 The Superposed State and Quantum Coherence**

In standard interpretations of quantum theory, the superposed state has ontological significance and cannot be interpreted "merely" as a probability of ignorance of some underlying deterministic process. Hence is not true that Schrödinger's cat is either alive or dead (it's just we don't know which). In the context of the Invariant Set Postulate, the superposed quantum state $\alpha|A\rangle + \beta|B\rangle$ has no fundamental ontological significance; it indeed describes a probability of ignorance, here ignorance of the intricate structure of the invariant set based on a sample space of trajectory segments in some neighbourhood on the invariant set. From this perspective, Schrödinger's cat is alive or dead, and not both.

**5.2 The Measurement Problem**

The Invariant Set Postulate provides a straightforward resolution of the measurement problem. By performing measurements we humans acquire information about the world state. That is, by empirical means, we become more knowledgeable about $I$ than could ever be gleaned from quantum theory alone. In quantum theory this empirical information is ingested through a "collapse of the wavefunction", a process external to quantum theory itself.

If the notion of superposition has no fundamental ontological significance, then neither has measurement. Rather, a measurement outcome "merely" labels a quasi-stationary quasi-stable region of the invariant set. There are no "jumps" in the world state as a result of measurement, since the world state was never in a superposition in the first place.

**5.3 The Copenhagen Interpretation vs Einstein Reality**

The Invariant Set Postulate reconciles Einstein's views on quantum theory, with those of Bohr, Heisenberg and Pauli, as summarised in the Copenhagen Interpretation.

On the one hand, consistent with Einstein's view, the Invariant Set Postulate indicates that quantum theory is incomplete in the sense that it is blind to the fractal structure of the invariant set and hence $D_I$. With respect to $D_I$, physics is both deterministic (no dice) and locally causal (no spooky effects).

On the other hand, the Invariant Set Postulate provides an objective basis for understanding why the observer is a partner in the very concept of reality.

From the Invariant Set Postulate, it is not meaningful to regard an individual quantum system as having any intrinsic properties independent of the invariant set on which the whole world state evolves. The invariant set is in part characterised by the experiments which inform us humans about it. Hence, the Invariant Set Postulate implies that it is not meaningful to regard a quantum sub-system as having any intrinsic properties independent of the measurements performed on the quantum system. Since experimenters play a role in determining the nature of these measurements, they manifestly also play a key role in defining the very concept of reality. This is one of the key tenets of the Copenhagen Interpretation.

## 5.4 Wave-Particle Duality, Bohmian Theory and Delayed-Choice Measurement

Following Feynman, wave-particle duality is often viewed as the central mystery of quantum theory. For example, how is it that a single particle "knows" how to avoid regions of destructive wave interference? In terms of the Invariant Set Postulate, the paradox is easily resolved, in principle at least. Since the state-space geometry of the invariant set is presumed to determine observed probabilities, then in a two-slit experiment, the world state where particles travel to regions of destructive wave interference will not lie on the invariant set, and therefore will not correspond to a state of physical reality. In this regard, one could regard the quantum potential of Bohmian theory (Bohm and Hiley, 2005) as a coarse-grain ("$L^2$") potential-well description of the constraint required to keep the world state on the invariant set.

Wave-particle duality is an example of the type of complementarity discussed above in the context of particle spin. Hence, having performed a "wave" experiment on a system, the corresponding counterfactual "particle" experiment (eg what would have happened if one slit were to have been closed off) lies off the invariant set and does not correspond to physical reality.

In this context, the Invariant Set Postulate can also readily account for the notion that the state of a quantum sub-system at $t = t_0$ can be influenced by measurements whose characteristics are seemingly only determined at $t_1 \gg t_0$. For example, Bob can postpone "deciding" whether to make a "wave" measurement or a "particle" measurement, long after the time when the quantum sub-system has to "make up its mind" whether to behave as a wave or as a particle. This apparent paradox is resolved by recalling that the notion of the invariant set is an atemporal one. Whether or not a world state lies on the invariant set at some time $t = t_0$ and hence is a point of reality, may depend on measurement events to the (indefinite) future $t_0$. (This is effectively another expression of the non-computability of the invariant set.) Labelling a trajectory "$A$=wave" at $t_0$ only makes sense if there is a corresponding quasi-stationary region "$A$=wave" of the invariant set into which the trajectory evolves.

**5.5 Nonlocality**

As Bell has pointed out (Bell, 1995), the notion that quantum mechanics is not locally causal (ie is nonlocal), depends on treating experimental parameters, such as the orientation of measuring devices, as free variables. Since $D_I$ is considered to be causal, the role of the Invariant Set Postulate as a restriction on the existence of free variables becomes a central issue in assessing whether quantum physics is nonlocal.

'tHooft (2007) relates the notion of a free variable to the Unrestricted Initial State condition, which he describes as having consequences similar to free will, but not clashing with determinism. In motivating the Unrestricted Initial State condition, 'tHooft says: "…we must demand that our model [of nature] gives credible scenarios for a universe for any choice of the initial conditions."

The Unrestricted Initial State condition is certainly plausible for any physical theory based solely on differential equations, eg the laws of classical physics. However, it manifestly fails in the context of Invariant Set Postulate. In this latter case, the only "credible scenarios" are associated with initial conditions which lie on the invariant set. As discussed above, this is a sufficient restriction to rule out counterfactual states.

Using the language of hidden variable theory, let $P(\lambda | \theta_a, \theta_b)$ denote the probability of some hidden-variable $\lambda$ given EPR measurement orientations $\theta_a$ and $\theta_b$. Conventionally, in hidden-variable theory, one assumes that $P(\lambda | \theta_a, \theta_b) = P(\lambda)$ (eg Weinstein, 2008) and this leads to the conventional Bell inequalities. By the discussion above, this condition is manifestly false in the context of the Invariant Set Postulate.

Some argue that the real mystery of quantum nonlocality is that Bell-inequality-violating correlations exist without any superluminal communication between particles. To explain this mystery, consider the following two "causal" statements: a) whenever I clap my hands in the college quad, I hear the echo a fraction of a second later; b) if I were to have clapped my hands in the college quad, I would have heard the echo a fraction of a second later. With respect to the Invariant Set Postulate, statement a) is true since the dynamics of acoustic waves is assumed causal. However, statement b) is not true since it is counterfactual. As such, the Invariant Set Postulate explains why it is possible for there to be no superluminal communication (arising from the truth of a)-type statements), whilst at the same time admitting Bell-inequality-violating correlations (possible because of the non-truth of b)-type statements).

Similarly one might ask why, if experimental settings at spacelike-separated locations cannot be varied independently of one another, should the outcome at one wing of an EPR experiment be independent of the experimental setting at the other? Like the notion of causality in the previous paragraph, here we note that the notion of "independent" can have two different meanings. Two timeseries can be viewed as statistically independent if they are uncorrelated, or more generally, if no functional dependence can be found between them. On the other hand, the same two timeseries can be viewed as dependent in the sense that any attempt to vary one timeseries keeping the other fixed will violate the conditions under which the timeseries have been constructed.

One philosophical objection to any restriction on the notion that experimental parameters are "free", is that one could set these experimental parameters on some whim, eg the toss of a coin, or the outcome of the Swiss Lottery, or the sixth digit of the frequency of a photon originating in Alpha Centuri. If measurement orientations can depend on outcomes which seem "irrelevant" as far as the evolution of the rest of the universe is concerned, then surely these experimental parameters are for all practical purposes, free? This was an argument John Bell himself used (Bell, 1995).

Here we refer back to the Takens Embedding Theorem - that the entire invariant set can be reconstructed from a sufficiently long timeseries of any component of the state vector, even if this component is energetically unimportant and seemingly irrelevant to the evolution of the rest of the universe. Hence it is inconsistent to conclude from the seemingly whimsical examples above, that measurement orientations can be made dependent on variables which are themselves irrelevant to or independent of the evolution of the rest of the world state.

Another potential objection to the notion that experimental parameters are other than "free", is that it might appear to make us humans, we who fix these parameters, seem no different to automata. However, by the Invariant Set Postulate, this is not the case. We humans are conscious beings. As such, we acknowledge as "real" the physical world around us. Hence, by the Invariant Set Postulate, we acknowledge the reality of $I$. That is, we acknowledge the reality of something which is fundamentally non-algorithmic. As stressed by Penrose (1989), no automaton would be capable of this!

Theories which constrain the existence or notion of free variables might be classed as "superdeterministic". However, it is important to distinguish the conventional idea of superdeterminism with that of the Invariant Set Postulate. Conventional superdeterminism is an *ad hoc* approach and hence unappealing. For example, superdeterminism presumes that the actual initial state of the universe is the only allowable initial state. Yet, why should this be? By contrast, the restrictions on experimental parameters implied by the Invariant Set Postulate are a consequence of a invoking a certain type of invariance - dynamical invariance. As mentioned above, invariance and symmetry are the bedrocks of theoretical physics and therefore not at all *ad hoc*. By contrast with standard superdeterministic thinking, the Invariant Set Postulate provides an appealing theoretical basis for constraining what would otherwise be unconstrained free variables.

In conclusion, insofar as the word "nonlocality" is shorthand for "not locally causal", then in the framework of the Invariant Set Postulate, quantum physics is not nonlocal.

**5.6 Emergence of Classicality**

It is well known that the classical limit cannot be reached by letting $\hbar \to 0$. This is consistent with the notion that the intertwined $(A, B)$ trajectory structures on $I$ persist to arbitrarily small scales in state space (cf Fig 1 in the vicinity of the fiducial trajectory). In this sense, the $\sqrt{\hbar}$ probability structure in the limit $\hbar = 0$ is singular (c.f. Berry, 2002).

As noted above, classical systems are not themselves bound by the Invariant Set Postulate. How then does the Invariant Set Postulate solve the "mystery" of emergence of classicality? One way to address this question is to consider the structure of the invariant set associated with time-averaged states. By the central limit theorem, we know that the measure of the invariant set associated with sufficiently long time-averaged states is Gaussian, and hence not fractal. A similar continuum measure would arise if $I$ was projected into a small dimensional subspace (cf "tracing out" the environmental degrees of freedom to use the language of decoherence theory). With such a smooth measure, the counterfactual set would not be the empty set, indeed in the (Gaussian) limit, a counterfactual set such as $\{\theta\}_C$ in Section 4.2 would be all of $0 \leq \theta \leq 2\pi$ and the (fractal-based) arguments about counterfactuality would fail. In this (time-averaged or projected) situation, the system would behave as if it were classical.

**5.7 Quantum Uncertainty**

Using the concept of algebraic abstraction discussed in Section 4.2, quantum theory provides a consistent mathematical definition of "state", irrespective of whether that quantum-theoretic state is related to a sample space of underlying deterministic processes or not. Quantum theory is blind to the intricate structure of the invariant set, and because the invariant set is not computable, quantum theory cannot be supplemented by any algorithm to determine whether the quantum-theoretic state is associated with an underlying sample space or not. In this sense, mathematical undecidability appears to be closely related to the notion uncertainty in quantum theory.

**5.8 Quantum Computing and the Multiverse**

The dynamical laws $D_I$ are presumed to encode the underlying geometry of $I$ in state space. What form should $D_I$ take? As discussed in Section 4, the symbolic form for $D_I$ on some fiducial trajectory segment, must reflect the symbolic properties of trajectories in a neighbourhood on the invariant set. In this respect, there is a partial analogy with the equation for a geodesic in space time, where the geodesic coordinates depend on the geometry of space time in the neighbourhood of the geodesic (as reflected in the appearance of connection coefficients in the geodesic equation). This required form for $D_I$ can be contrasted with dynamical evolution equations $\dot{\mathbf{X}} = \mathbf{F}(\mathbf{X})$ in classical physics, where time derivatives are determined solely by the values of the state $\mathbf{X}$ at any chosen point in state space, and not on surrounding values. This may provide the conceptual basis to understand the relative power of quantum computing over classical computing.

From a geometric perspective, the notion that the quantum theoretic state is a "coarse-grain" probabilistic representation based on a sample space from neighbourhoods of the invariant set, seems comprehensible. On the other hand, if looked at from a perspective where attention is focussed exclusively on dynamical evolution of the fiducial trajectory, such a probabilistic notion may seem bizarre: it may take such a long time (ie such a large parameter length $t$) before the world state returns to a $\sqrt{\hbar}$ neighbourhood of its current state on the invariant set, that two neighbouring world states actually belong to different eons of the universe (eg Bojowald, 2007; Penrose, 2008). Forming probabilities from such a multiverse sample space may seem almost

Everettian in concept, and very difficult to comprehend. This is why the atemporal geometric picture of the invariant set, proposed here, has appeal.

**5.9 The Role of Gravity in Quantum Physics**

Gravity has often been suggested as playing a role in quantum theory, principally as a mechanism that induces quantum state vector collapse (cf Objective Reduction; Penrose, 1989). However, at an ontological level, the Invariant Set Postulate does not require superposed states and hence does not require a collapse mechanism, gravitational or otherwise.

On the other hand, the order-of-magnitude estimates provided by Penrose (1989), that gravitational processes can be locally significant when a quantum sub-system and a measuring apparatus interact, seem persuasive. Here we would interpret these estimates as supporting the notion that gravity plays a key role in defining the state space geometry of the invariant set, in particular in defining the regions of relative stability (small local Lyapunov exponents) and relative instability of $I$ (large local Lyapunov exponents). As discussed above, black hole thermodynamics may additionally provide the mechanism which leads to the dimensional reduction of the invariant set compared with that of the underlying state space.

Indeed this leads to the following rather radical suggestion. If the geometry of $I$ is to be considered primitive, then the geometric properties of the invariant set which lead to certain regions being relatively stable and other regions unstable, should be considered a generalisation of the notion introduced by Einstein, that the phenomenon we call "gravity" is merely a manifestation of some more primitive notion of geometry - here the geometry of a dynamically-invariant subset of state space. As such, a challenge for the future will be to try to unify the notions of pseudo-Riemannian geometry for space-time, and fractal geometry for state space. This is a very different perspective on "quantum gravity" than suggested by any existing approaches to the subject.

From this we can make two gravitationally relevant predictions. Firstly, since gravitational processes are not needed to collapse the quantum state vector, experiments to detect gravitational decoherence (eg as would be needed in the Objective Reduction mechanism) may fail. By contrast with Objective Reduction, $I$ could be seen as providing the preferred basis, with respect to which conventional non-gravitational decoherent processes operate. Secondly, if gravity should be seen as a manifestation of the heterogeneity in the geometry of the invariant set, then attempts to quantise gravity with the framework of standard quantum theory, will also fail. As such, it is misguided to assume (as almost all serious attempts have so far done) that "theories of everything", can be formulated within conventional quantum theory.

## 6 Conclusions

Principles of invariance and symmetry form the bedrock of physics. Based on a type of dynamical invariance, a new law of physics is proposed, directly relevant at the cosmological level. Specifically, the Invariant Set Postulate subordinates the notion of the differential equation and elevates as primitive, a dynamically invariant fractal geometry in the state space of the universe. This geometry is used to define the notion of physical reality - states of physical reality are precisely those on the invariant set. It is suggested that this postulate has profound implications for our understanding of quantum physics and the corresponding role of gravity.

The Invariant Set Postulate is motivated by two quite disparate ideas in physics. Firstly, certain nonlinear dynamical systems have measure-zero, nowhere-dense, self-similar non-computational invariant sets. Secondly, the behaviour of extreme gravitationally bound systems is described by the irreversible laws of thermodynamics at a fundamental rather than phenomenological level.

In the 1960s, the introduction of global space-time geometric and topological methods, transformed our understanding of classical gravitational physics (Penrose, 1965). It is proposed that the introduction of global geometric and topological methods in state space, may similarly transform our understanding of quantum gravitational physics. Combining these rather different forms of geometry may provide the missing element needed to advance the search for a unified theory of physics.

Future papers will attempt to provide the mathematical detail required to develop this exploratory analysis into a rigorous physical theory.